\documentclass[]{aa}
\usepackage{graphicx}

\renewcommand{\phi}{\varphi}
\begin{document}

\title{Stability of Terrestrial Planets in the Habitable Zone of Gl~777~A, HD 72659, Gl 614, 47 Uma and HD 4208}

\titlerunning{Earthlike planets}

\author{N.~Asghari\inst{1}, C.~Broeg\inst{1}, L.~Carone\inst{1},
R.~Casas-Miranda\inst{1}, J.C.~Castro
  Palacio\inst{1}, I.~Csillik\inst{1}, R.~Dvorak\inst{2},
  F.~Freistetter\inst{2}, G.~Hadjivantsides\inst{1}, H.~Hussmann\inst{1},
A.~Khramova\inst{1},
  M.~Khristoforova\inst{1}, I.~Khromova\inst{1}, I.~Kitiashivilli\inst{1},
  S.~Kozlowski\inst{1}, T.~Laakso\inst{1},
  T.~Laczkowski\inst{1}, D.~Lytvinenko\inst{1}, O.~Miloni\inst{1},
R.~Morishima\inst{1}, A.~Moro-Martin\inst{1},
  V.~Paksyutov\inst{1}, A.~Pal\inst{1}, V.~Patidar\inst{1},
B.~Pe\v{c}nik\inst{1},
  O.~Peles\inst{1}, J.~Pyo\inst{1}, T.~Quinn\inst{5},
A.~Rodriguez\inst{1}, C.~Romano\inst{1}\fnmsep\inst{3},
  E.~Saikia\inst{1}, J.~Stadel\inst{4}, M.~Thiel\inst{3},
  N.~Todorovic\inst{1}, D.~Veras\inst{1}, E.~Vieira Neto\inst{1},
  J.~Vilagi\inst{1}, W.~von Bloh\inst{1}, R.~Zechner\inst{1}, \and
  E.~Zhuchkova\inst{1}
}
\authorrunning{Asghari et al.}

\offprints{R.\ Dvorak, \email dvorak@astro.univie.ac.at}

\institute{
Helmholtz-Institute for Supercomputational Physics, Am Neuen Palais 10,
D-14469 Potsdam, Germany
\and
Universit\"atssternwarte Wien, T\"urkenschanzstr. 17, A-1180 Wien, Austria
\and
Institut f\"ur Physik, Universit\"at Potsdam, Am Neuen Palais 10, D-14469
Potsdam,
  Germany
\and
Institut f\"ur Theoretische Physik, Universit\"at Z\"urich,
Winterthurerstr. 190, 8057--Z\"urich, Switzerland
\and
Department of Astronomy, University of Washington, Seattle 98195-1580,
USA}

\date{Received; accepted}

\abstract{
We have undertaken a thorough dynamical investigation of five extrasolar
planetary systems using extensive numerical experiments on the supercomputer of
the Max Planck Institute for Gravitational Physics (Albert Einstein Institute).
This work was performed as part of the Helmholtz Institute for Supercomputational
Physics Summer School on ``Chaos and Stability in Planetary Systems'' (2003).
The systems Gl~777~A, HD~72659, Gl~614, 47~Uma and HD~4208 were examined
concerning the question of whether they could host terrestrial like planets in
their habitable zones (=HZ).

First we investigated the mean motion resonances between 
fictitious terrestrial planets and the existing gas giants in these five
extrasolar systems. Then a fine grid of initial conditions for a potential 
terrestrial planet within the HZ was chosen for each system, from which the stability of orbits
was then assessed by direct integrations over a time interval of 1 million years.
For each of the five systems the 2-dimensional grid of initial conditions contained 
80 eccentricity points for the Jovian planet and up to 160 semimajor axis points 
for the fictitious planet.
The computations were carried out using a Lie-series integration method with an 
adaptive step size control. This integration method achieves machine precision 
accuracy in a highly efficient and robust way, requiring no special adjustments
when the orbits have large eccentricities. 

The stability of orbits was examined with a determination of the R\'enyi entropy, estimated from
recurrence plots, and with a more straight forward method based on the maximum eccentricity
achieved by the planet over the 1 million year integration.
Additionally, the eccentricity is an indication of the habitability of a
terrestrial planet in the HZ; any value of $e>0.2$ produces a significant temperature difference
on a planet's surface between apoapse and periapse.
The results for possible stable orbits for terrestrial planets in habitable
zones for the five systems are summarized as follows:
for Gl~777~A nearly the entire HZ is stable, for 47~Uma, HD~72659 and HD~4208 terrestrial planets can 
survive for a sufficiently long time, while for Gl~614 our results exclude terrestrial planets moving 
in stable orbits within the HZ.

Studies such as this one are of primary interest to future space missions 
dedicated to finding habitable terrestrial planets in other stellar systems.
Assessing the likelihood of other habitable planets, and more generally the 
possibility of other life, is the central question of astrobiology today.
Our investigation indicates that, from the dynamical point of view, 
habitable terrestrial planets seem to be quite compatible with many of the currently 
discovered extrasolar systems.
\vspace{2cm}

 }

\maketitle

\section{Introduction}

In this study\footnote{The results presented here were derived during the
$3^{\mathrm{rd}}$ Helmholtz Summerschool at the Helmholtz Institute for
Supercomputational Physics
 in Potsdam: ''Chaos and Stability in Planetary Systems'' from September 
$1^{\mathrm{st}}$ to September $26^{\mathrm{th}}$ 2003} 
the extrasolar systems Gl~777~A, HD~72659, Gl~614, 47~Uma and HD~4208 were examined
on the question of whether they could host additional terrestrial like planets in 
their habitable zones (=HZ). 

Since the discovery of the first extrasolar planetary system about 10
years ago~(Mayor \& Queloz \cite{mayor}), a major point of dynamical investigations 
has been the determination of stable regions in extrasolar planetary systems, where
additional planets on stable orbits could exist. Today we have evidence for 
105 planetary systems with 120 planets, where 13 systems have more than one
planet. 

Because of the observational methods which are in use, our knowledge of extrasolar planets is
highly biased: almost all planets were detected via ground based measurements
of the central star's radial velocity. This method favors the detection of
very large (Jupiter-sized) planets that move very close to the main
star. Several space missions are planned (i.e. {\sf COROT}, {\sf KEPLER}, 
{\sf TPF}, {\sf DARWIN}) that will be able to find planets with a much smaller 
mass (comparable to that of the Earth) by using other detection methods. 
The first results of these missions will only be available after more 
than a decade, however, theoretical studies can currently make predictions on the 
existence of stable terrestial planets within the known systems.
This serves to establish testable theories of planet formation and evolution,
as well as aiding the future searches for such habitable terrestial planets.

When one looks at the distribution of semimajor-axes of all known exosolar
planets, one can see that they are located from $\sim 0.02$ AU to $\sim 6.5$ AU.
These distances cover the so-called ``habitable zone'' (=HZ)\footnote{The
  region were a planet fulfills all the requirements of being able to
  develop life and where liquid water can 
be present on the surface of the planet.} of a typical main sequence star, 
ranging from $\sim 0.7$~AU to $\sim 1.5$ AU~(Kasting et al. \cite{Kasting93}). 
Some of these large planets move in the region around 1 AU that overlaps with the HZ, 
and leave no room for possible terrestrial planets orbiting the host star within the HZ
-- only a large satellite (like Titan) or a Trojan like planet could have all
the properties which are necessary for orbital stability and also the
development of a stable atmosphere in such cases.
When the large planets move well outside (or inside) the HZ they will perturb any 
additional undetected planet that may move within this zone.

For extrasolar systems with only one planet the possible existence of additional planets 
has been investigated (e.g. Dvorak et al. \cite{Dvorak03b}; P\'al \& S\'andor~\cite{Pal03}).
In an extensive paper (Menou \& Tabachnik~\cite{Menou03}) the stability of possible 
terrestrial planets was studied for all extrasolar systems and a classification according to 
their stability was established, but there the role of resonances was underestimated.
Resonances -- mean--motion resonances and secular resonances --  can either
stabilize the motion of a planet and protect it from close encounters and
collisions or, in other cases, intensify the gravitational perturbations and
therefore destroy the stability of the orbits. 
Resonances between the orbits of the observed large planets 
have been studied
 (Callegari et al.~\cite{Callegari02}; Hadjidemetriou~\cite{Hadjidemetriou02};
 \'Erdi \& P\'al~\cite{Erdi03}), but a detailed study of  the  
perturbations on possible additional planets due to resonances is still
lacking. Thus, any study of the dynamics
inside the HZ of an extrasolar system has also to include an investigation of
the resonances.

Determinations of the dynamical stability of multiple planetary
systems (Kiseleva et al. \cite{Kiseleva02a},~\cite{Kiseleva02b}; 
Beaug\'e \& Michtchenko~\cite{Beauge03}) and of planets in double stars
(Pilat-Lohinger \& Dvorak \cite{Pilat02a}; Pilat-Lohinger et al.~\cite{Pilat02b},~\cite{Pilat03};
Dvorak et al.~\cite{Dvorak03a}; Lammer et al.~\cite{Lammer03}) have been done
recently.

In this article we  investigate the dynamics inside the habitable zone of five
different extrasolar systems: Gl~777~A, 47~Uma, HD~72659, Gl~614 and
HD~4208. Except 47~Uma, all these systems consist of the central star and a
Jovian planet. According to the spectral type of the main star, the
habitable zone was determined; a detailed analysis of the mean motion resonances
inside the HZ was carried out and the full width of the HZ was investigated
according to the stability of possible additional terrestrial planets.
This article is organized as follows: Section~\ref{sec2} describes the five 
extrasolar systems for which we performed a dynamical study for additional planets, 
in section~\ref{sec3} we introduce the dynamical models and the methods used to 
analyze the time series data resulting from our numerical experiments.
Section~\ref{sec4} shows how we dealt with the terrestrial planet orbits 
in the major mean--motion resonances of the Jovian planet(s). 
Subsequent sections focus on each of the 5 systems individually, detailing 
our investigations and results.
We conclude with a summary of these results and their implications, including 
comparison to earlier, less detailed studies by Menou \& Tabachnik~(\cite{Menou03}). 
The appendicies include concise descriptions of the analytical and numerical methods used, 
and may serve as a useful reference to the reader.

\section{Description of the Extrasolar Systems}
\label{sec2}

Table~\ref{tab1} shows the main characteristics of the five extrasolar planetary
systems which we investigated in this study.
Except the system 47~Uma (which hosts two planets), all consist of a
giant planet moving outside the HZ of the star; this is a situation
similar to our Solar System (=SS) when we take into account only Jupiter. 
Thus, an appropriate dynamical model is the restricted three body problem (star +
planet + massless terrestrial planet). A different dynamical model was taken for the 
system 47~Uma, which looks quite similar to our SS when only Jupiter
and Saturn are considered. Both 47~Uma and the SS have the more massive planet
orbiting closer to the star with the second planet at 
approximately twice the distance. Furthermore, the 47~Uma planets have 
roughly a 3:1 mass ratio and exhibit low eccentricity orbits very much like 
Jupiter and Saturn in our own SS.
The restricted four body problem, known to be a good model for an asteroid in the 
SS (taken as a massless body), additionally considers the secular resonances acting 
on the motion of a massless test body and will also serve as a good model for 
terrestrial planets within 47~Uma.

\begin{table}
      \caption[]{Parameters of the investigated systems -- obtained from {\tt
          http:{\//}{\//}www.obspm.fr{\//}planets} (1. Sep. 2003)}
         \label{tab1} 
         \begin{tabular}{llll}
            \hline
            \noalign{\smallskip}
            Name & Mass & Semimajor& Eccentricity  \\
                 &      &  axis [AU]& \\
            \noalign{\smallskip}
            \hline
            \noalign{\smallskip}
            Gl 777 A (G6IV)& $0.90$ ${\mathrm{M_\odot}}$& -- & --\\
            Gl 777 A b& $1.33$ ${\mathrm{M_{J}}}$& $4.8$ & $0.48 \pm 0.2$\\
            \hline
            HD 72659 (G0V) & $0.95$ ${\mathrm{M_\odot}}$& -- & --\\
            HD 72659 b & $2.55$ ${\mathrm{M_{J}}}$& $3.24$ & $0.18$\\
            \hline
            Gl 614 (K0V) & $1.00$ ${\mathrm{M_\odot}}$& -- & --\\
            Gl 614 b & $4.89$ ${\mathrm{M_{J}}}$& $2.85$ & $0.38$\\
            \hline
            HD 4208 (G5V) & $0.93$ ${\mathrm{M_\odot}}$& -- & --\\
            HD 4208 b & $0.80$ ${\mathrm{M_{J}}}$& $1.67$ & $0.05$\\
            \hline
            47 Uma (G0V) & $1.03$ ${\mathrm{M_\odot}}$& -- & --\\
            47 Uma b & $2.54$ ${\mathrm{M_{J}}}$& $2.09$ & $0.061 \pm 0.014$\\
            47 Uma c & $0.76$ ${\mathrm{M_{J}}}$& $3.73$ & $0.1 \pm 0.1$\\
            \noalign{\smallskip}
            \hline
         \end{tabular}
   \end{table}

In a wider sense we can say that all the central stars are solar like stars
with masses slightly smaller or larger than the Sun. The spectral types for
these main sequence stars (G0 to K0) allow HZs in the range
between 0.5 AU and 1.5 AU. With the exception of HD 4208 and 47~Uma all the
Jovian planets' orbits have significantly larger eccentricities than that of
Jupiter with sometimes very large uncertainties. 
On the other hand, we expect the determined semimajor axes to be rather precise. 
Our parameter study will take into account this discrepancy in the precision of 
the measured orbital elements of the exoplanets under study and will be detailed
in the following section.
The minimum masses of the Jovian planets lie exactly in the range we expect for 
gas giants. With regard to the unknown inclinations of these extrasolar systems, 
we always took the masses of table~\ref{tab1} as correct, assuming that the orbital 
plane of the system is seen edge--on. It should be kept in mind that the true masses 
of these planets could be significantly larger.
Statistically we can say that in 5 of 6  cases the inclination will
change the mass by a factor smaller than four compared to the one given in the
tables. Recent computations have shown that with these larger masses the
strenghts of the perturbations will not significantly change
(Sandor et al.~\cite{zsandor04}).

\section{Simulation Method and Stability Analysis}
\label{sec3}

The availability of a supercomputer with 128 processors\footnote{The PEYOTE cluster at 
the Max Planck Institute for Gravitational Physics (Albert Einstein Institute): 
www.aei-potsdam.mpg.de/facilities/public/computers.html} for this investigation
enabled the direct computation of orbits to assess stability. Futhermore, the
extent of the computational resources favoured the use of a very precise numerical
integration scheme, the Lie-integration method, which is free from numerical difficulties
experienced by other (lower order) techniques, particularly in the case of highly
eccentric orbits. The Lie-integration method uses an adaptive stepsize and is quite 
precise and fast, as has been shown in many comparative test computations with other
integrators such as Runge-Kutta, Bulirsch-Stoer or symplectic integrators. Although,
symplectic integrators are very effective when eccentricities remain small, the
Lie integrator is the better choice in studies such as this one, where very large
eccentricity orbits are explored.
We detail this integration method in Appendix~\ref{ApB}, and further details can 
be found in Hanslmeier \& Dvorak~\cite{Hanslmeier84} as well as 
Lichtenegger~\cite{Lichtenegger84}.

The uncertainties in the orbital elements derived from observations make it
sufficient to consider only Newtonian forces in a dynamical model of two 
massive bodies (the central star and the discovered planet) and a massless 
test body, representing a potential terrestrial planet. As mentioned previously, 
this system corresponds to the three dimensional restricted 3--body problem.
For the 47~Uma system where two Jovian planets are known, we followed the
dynamics of a restricted 4--body problem with a central star, both discovered 
Jovian planets, and a massless test body representing a potential terrestrial 
planet. This increased the computing time considerably for 47~Uma compared to 
the other restricted 3--body systems.

For the integrations we use a fine 2--dimensional grid of initial
conditions; for the terrestrial planets varying the distance to the
central star, exploring the HZ, and for the (inner -- 47~Uma) Jovian planet varying the 
eccentricity of its orbit. We vary the eccentricity of the Jovian planet because 
the values derived from observations have large uncertainties
\footnote{In most cases they can only be determined from the shape of the 
periodic stellar velocity curve, while $M \sin{i}$ and the semimajor axis
come from the zero'th order characteristics, the amplitude and period of 
this curve respectively.}, and yet this is an important parameter 
for stability of other planets within the system. Table~\ref{tab:2} outlines the
ranges for our initial conditions.
We always started the initial orbits of the terrestrial planets as being circular; 
they turned out to deform quite quickly, exploring a range of more eccentric orbits so 
that varying their initial eccentricities would seem to be extraneous. 
Besides the eccentricity of the fictitious planet's orbit, guided by the 
formation scenario for rocky terrestrial planets (Richardson et al.~\cite{Richardson00}) the inclination was set to almost zero. 
As a final detail we mention that the integration of the terrestrial planet's orbit 
was always started on the connecting line between the star and the Jovian planet. 
An integration time of 1 million years was chosen, which is long enough to 
unveil the stability character of the planets for the 5 systems of this study. 
The default parameter grid of 80 by 80 which we used for the 5 systems leads to a 
grand total integration time of several $10^{\mathrm{10}}$ years, when we also 
include test computations and the fact that some of the orbits were integrated for an 
extended $10^{\mathrm{7}}$ years.

\begin{table}
      \caption[]{Initial semimajor axes of the fictitious terrestrial planets
        $a_{\mathrm{T}}$ and the eccentricities of the jovian planets  $e_{\mathrm{G}}$}
         \label{tab:2} 
         \begin{tabular}{lrl}
            \hline
            \noalign{\smallskip}
            System & $a_{\mathrm{T}}$ [AU]& $e_{\mathrm{G}}$ \\
            \noalign{\smallskip}
            \hline
            \noalign{\smallskip}
            Gl 777 A & $0.5-1.3$ & $0.4-0.5$ \\
            47 Uma   & $0.5-1.5$ & $0.0-0.12$ \\
            HD 72659 & $0.4-1.6$ & $0.08-0.3$ \\
            Gl 614   & $0.5-1.5$ & $0.3-0.5$ \\
            HD 4208  & $0.55-1.4$ & $0.0-0.2$ \\
            
            \noalign{\smallskip}
            \hline
         \end{tabular}
   \end{table}

For the analysis of stability we used -- on one hand -- a straightforward check
based on the eccentricities. For this we examined the behavior of the eccentricity of
the terrestrial planets along their orbit and used the {\em largest} value as 
a stability criterion; in the following we call it the maximum eccentricity method (=MEM).  
This simple 
check has already been used in other studies of this kind and was found to be quite
a good indicator of the stability character of an orbit (Dvorak et al.~\cite{Dvorak03a}). 
An orbit was deemed unstable when the eccentricity exceeded a value of
$e=0.5$, after which we stopped further computation. In all our former studies this
stability limit turned out to be an appropriate tool because all the
terrestrial planet orbits with $e=0.5$ turned out to suffer, sooner or later, from a 
close encounter with the large planet, 
causing the terrestrial planet to escape (Dvorak et al.~\cite{Dvorak03a}).
For the habitability of a planet we also used an additional criterion based
directly on the eccentricity of the orbit within the HZ. This was done in
order to take into account the variations in the ``solar'' insolation on the
surface of the terrestrial planet. To good approximation (Lammer 2004;
private communication), requiring $e<0.2$ is sufficient to keep this
variation in insolation small enough during an orbit.

On the other hand we computed the R\'enyi entropy, a measure which is often
used in nonlinear dynamics to determine how predictable an orbit is. We
estimated the R\'enyi entropy by means of Recurrence Plots (RPs), a tool of
data analysis that has found numerous applications in many different fields in
the last years (Webber et al.~\cite{Webber}; Marwan et al.~\cite{norbert_geo}; 
Thiel et al.~\cite{thiel_and}). RPs were
initially introduced to simply visualize the behaviour of trajectories in
phase space. The distances of every pair of points of the system's
trajectory are represented in a 2-dimensional binary
matrix. In this way RPs yield different patterns depending on the
system's character. One can introduce different measures that quantify
the obtained structures. It was shown in Thiel et al.~(\cite{thiel_and}), that it is
also possible to estimate the R\'enyi entropy based on the distribution of
diagonal lines obtained in the RP.
The details of this second approach are covered in Appendix~\ref{ApA}.

\section{The Stability within Resonances}
\label{sec4}

For the investigation of the resonances, we choose initial conditions placed in the most 
relevant mean--motion resonances (=MMRs) of the fictitious planet with the Jovian planet inside
but also outside the HZ. 
These resonances were checked for stability in 8 different positions of the terrestrial planet 
(corresponding to $M=0^{\mathrm{\circ}},45^{\mathrm{\circ}},90^{\mathrm{\circ}},
135^{\mathrm{\circ}},180^{\mathrm{\circ}},225^{\mathrm{\circ}},270^{\mathrm{\circ}},315^{\mathrm{\circ}})$.  
Additionally the computations were carried out with the Jovian planet initially placed at
the apoastron and periastron. For a detailed list of the resonant positions that were
investigated for every system see table~\ref{tab3}.

\begin{table*}
\begin{minipage}{\linewidth}
\renewcommand{\footnoterule}{}
\begin{center}
      \caption[]{Stability of orbits in mean motion resonances. The numbers
        give the stable orbits according to the 8 different initial conditions}
         \label{tab3} 
         \begin{tabular}{lcccccccccc}
            \hline
            \noalign{\smallskip}
&\multicolumn{2}{c}{Gl 777 A}&\multicolumn{2}{c}{47 Uma\footnote{Note that in
            the case of the 47 Uma system, 
            where two
            Jovian planets are known, we did not use peri- and apoastron
            position as initial conditions, but 2 different modes
            corresponding to an aligned or anti-aligned configuration of the
            two major bodies.}}&\multicolumn{2}{c}{HD
            72659}&\multicolumn{2}{c}{Gl 614\footnote{Besides the 7 given
            resonances, we calculated the motion inside the 7:2, 9:2 and 8:3
            resonance -- again, we only found unstable
            motion.}}&\multicolumn{2}{c}{HD 4208\footnote{For this system, we
            calculated all resonances up to the $4^{\mathrm{th}}$ order (see
            section 4); with the exceptions of the 15:11 and the 13:9 MMRs the other resonant positions showed
            the same amount of predominantly stable motion.}}\\
\hline    
\noalign{\smallskip}
& P & A & Mode I & Mode II & P  & A  & P  & A  & P & A\\
\hline    
\noalign{\smallskip}
5:1& 1 & 0 & 8  & 6   & 8  & 8  & 0  & 0  & 8 & 8 \\
4:1& 0 & 0 & 3  & 0   & 8  & 8  & 0  & 0  & 8 & 8 \\
3:1& 0 & 0 & 0  & 0   & 4  & 3  & 0  & 0  & 8 & 7 \\
5:2& 0 & 0 & 6  & 7   & 2  & 2  & 0  & 0  & 3 & 2 \\
7:3& 0 & 0 & 1  & 1   & 1  & 0  & 0  & 0  & 8 & 8 \\
2:1& 0 & 0 & 1  & 2   & 3  & 3  & 0  & 0  & 8 & 8 \\
5:3& 0 & 0 & -- & --  & 2  & 1  & -- & -- & 7 & 5 \\
3:2& 2 & 1 & -- & --  & 0  & 2  & 0  & 0  & 7 & 8 \\
4:3& 0 & 0 & 4  & 0   & -- & -- & -- & -- & 3 & 1 \\
\hline
\noalign{\smallskip}
Sum [\%]& 4.7 & 1.6 & 28.6 & 69.6& 42.2 & 43.8 & 0.0 & 0.0 & 83.3 & 76.4 \\
Total Sum [\%] & \multicolumn{2}{c}{3.1} & \multicolumn{2}{c}{34.9} & \multicolumn{2}{c}{43.0} & \multicolumn{2}{c}{0.0} & \multicolumn{2}{c}{79.9}\\
           
            \hline
         \end{tabular}
\end{center}

\end{minipage}
\end{table*}

\begin{figure}[hhh]
\begin{center}
\includegraphics[width=3.4in]{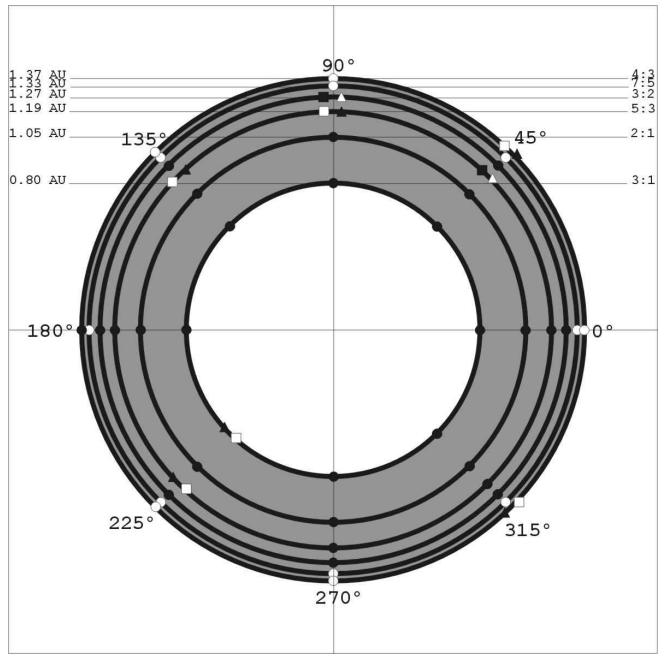}
\caption{Schematic view of the stability of orbits in the resonances of
$1^{\mathrm{st}}$ and $2^{\mathrm{nd}}$ order in HD 4208. Full (empty) circles stand for stable
(unstable) orbits in apoastron and periastron position. When the stability is
different we mark the apoastron by a triangle, the periastron with a square.}
\label{fig1}
\end{center}
\end{figure}

\begin{figure}[hhh]
\begin{center}
\includegraphics[width=3.4in]{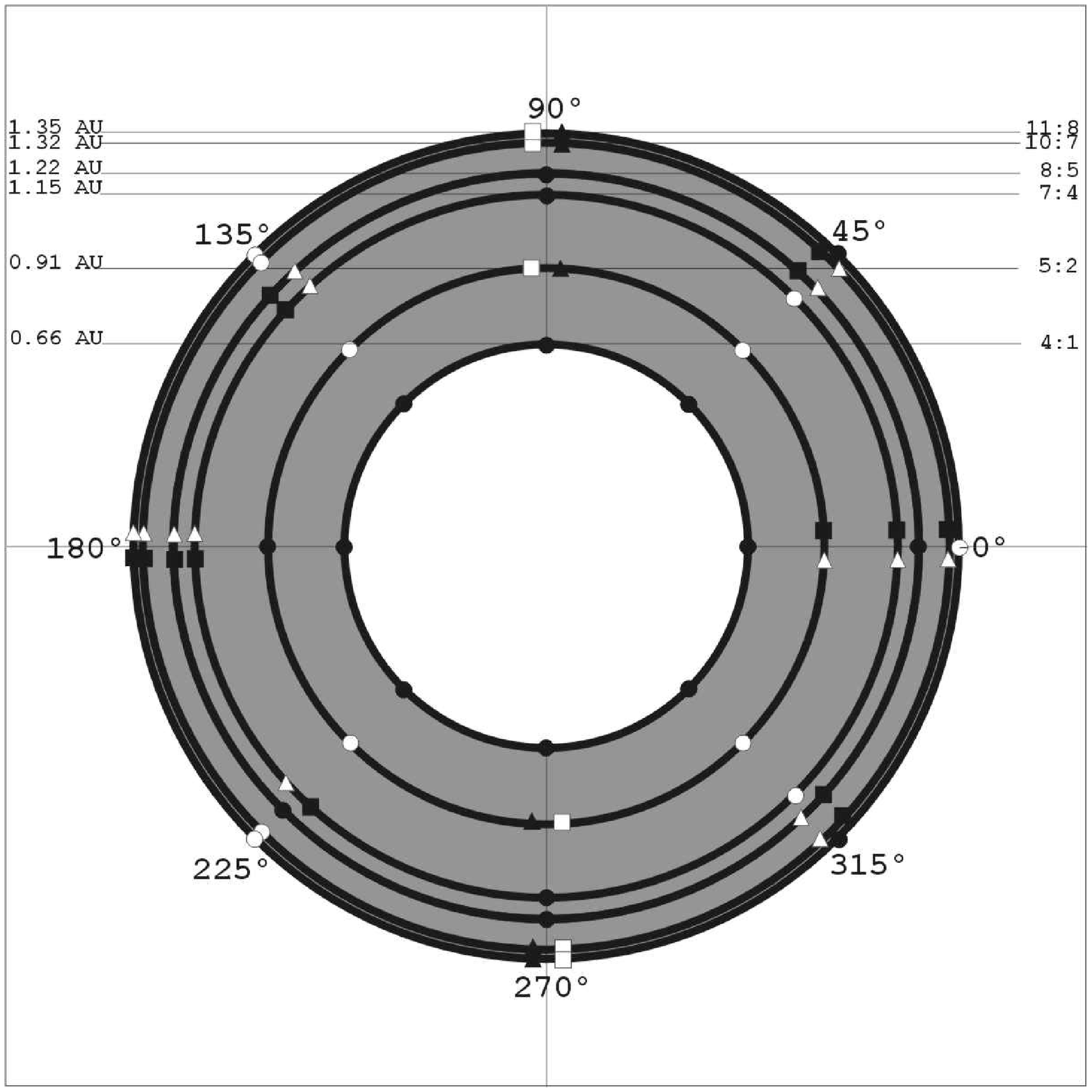}
\caption{Schematic view of the stability of orbits in the resonances of
 $3^{\mathrm{rd}}$ order in HD 4208. Description like in figure~\ref{fig1}.} 
\label{fig2}
\end{center}
\end{figure}

\begin{figure}[hhh]
\begin{center}
\includegraphics[width=3.4in]{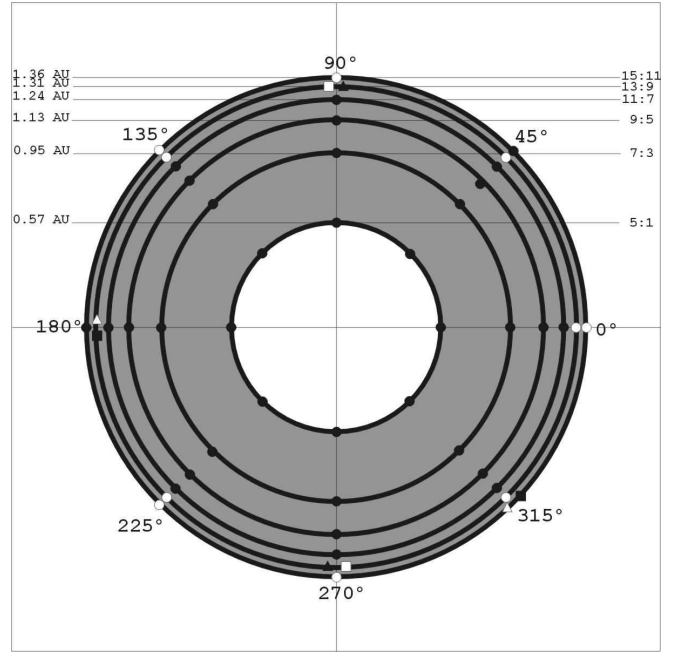}
\caption{Schematic view of the stability of orbits in the resonances of
 $4^{\mathrm{th}}$ order in HD 4208. Description like in figure~\ref{fig1}.}
\label{fig3}
\end{center}
\end{figure}

As an example we discuss the results of the investigation of
the MMRs for the system HD~4208. We studied the
following mean motion resonances up to the fourth order:
2:1, 3:2 and 4:3 (first order); 3:1, 5:3, 7:5 (second order); 4:1, 5:2, 7:4,
8:5, 10:7 and 11:8 (third order); 
5:1, 7:3, 11:7, 13:9, 15:11 (fourth order).
As shown in the picture for the first set of resonances ($1^{\mathrm{st}}$ and $2^{\mathrm{nd}}$
order) the orbits close to the central star, which move well inside the HZ,
are all stable (figure~\ref{fig1}). For the MMRs close to the Jovian planet 
we can see a preference for stable orbits for the
initial conditions $M=0^\mathrm{\circ}$, and $180^\mathrm{\circ}$. For the $3^{\mathrm{rd}}$ order resonance
(figure~\ref{fig2}) the picture is very inhomogeneous; what we can see is, that for the
Jovian planet in the apoastron position the orbits are stable even for the
resonances close to the giant planet.  The $4^{\mathrm{th}}$ order resonances are not
destabilizing an orbit as we can see from figure~\ref{fig3}; most of them are stable! 
The percentage of stable orbits in resonances is very large for HD~4208.

Details of the results from the investigation of resonances for all systems
can be found in table~\ref{tab3}.
For the resonances acting in three of the systems we can say that far from the 
perturbing planet almost all of them are stable; closer to the perturbing planet, 
they are more and more unstable (47~Uma, HD~72659 and HD~4208). Two systems
are very   much dominated by unstable motion in resonances: Gl~777~A and Gl~614.

\section{Gl 777 A}

The first discovery of a planet in Gl~777~A (=HD~190360) was reported by Naef
et al.~(\cite{Naef03}) from the Geneva group of observers. This extrasolar planetary
system is a wide binary with a very large separation (3000 AU);
for our dynamical investigations of motions close to one star there was no need
to take into account the perturbations of the very far companion. 
The central star is of spectral type G6 IV with $0.9 M_{\odot}$ and has
a planet of minimum mass 1.33~$M_{\mathrm{J}}$ with a semimajor-axis 
of 4.8~AU. 
Because of the large eccentricity (e=0.48) the possible region of
motion for additional planets is confined to $a<2.4$ AU (= periastron). 
Nevertheless, to have a global stability picture of possible additional planets, 
we investigated the stability in the region of the MMR
from the 4:3 to the 5:1 resonance located at a=1.64 AU. From table~\ref{tab3} one can 
see that only a few percent of the orbits started in the MMRs are stable.

\begin{figure}[hhh]
\begin{center}
\includegraphics[width=3.4in]{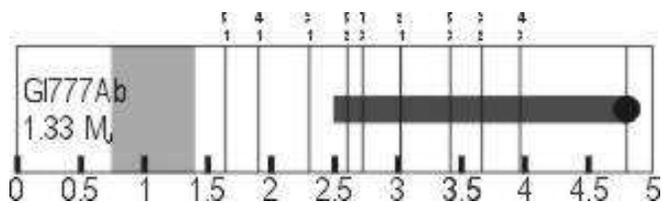}
\caption{Main characteristics of the extrasolar system Gl~777~A. The light grey
  region shows approximately the position of the HZ; the dark grey bar
  indicates how closely the planet approaches the central star in its orbit.}
\label{fig4}
\end{center}
\end{figure}

\begin{figure}[hhh]
\begin{center}
\includegraphics[width=3.4in]{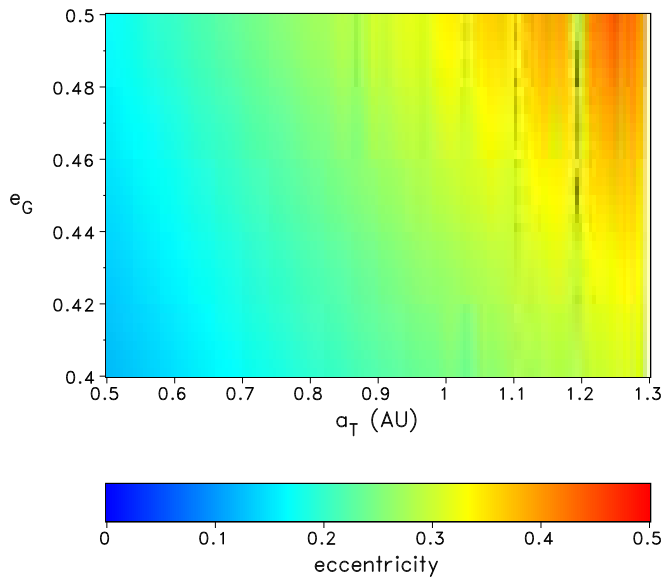}
\caption{
Initial condition diagram for fictitious planets in the system Gl~777~A: initial semimajor
  axes of the planet versus the eccentricity of the Jovian planet. The maximum
  eccentricity of an orbit during its dynamical evolution is marked with different colors.}
\label{fig5}
\end{center}
\end{figure}

\begin{figure}[hhh]
\begin{center}
\includegraphics[width=3.4in]{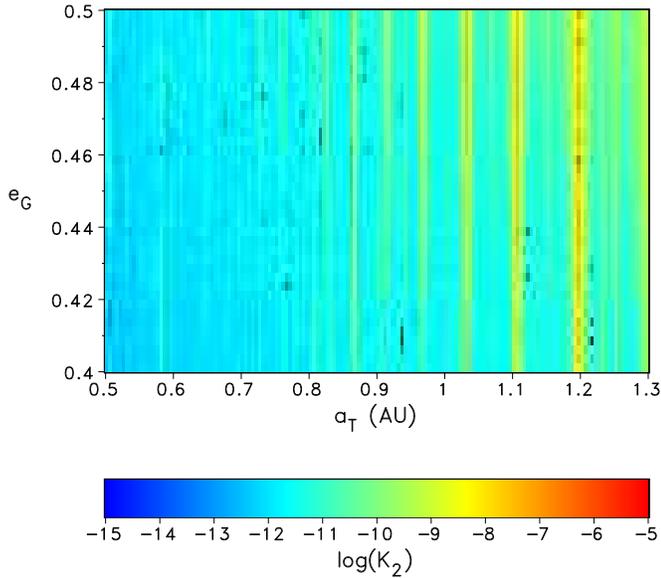}
\caption{Initial condition diagram for fictitious planets in the system Gl~777~A: initial 
semimajor
  axes of the planet versus the eccentricity of the Jovian planet. The value
  of the entropy (=entropy plot) of an orbit is marked in different colors.}
\label{fig6}
\end{center}
\end{figure}

The interesting region of habitability (see figure~\ref{fig4}), where planets could have temperature
conditions to
allow  liquid water on the surface, corresponds roughly to $0.7< a < 1.3$ AU, where we
ignore the eccentricity of the terrestrial planet. 
We have started our
computations in a larger region ($0.5 < a < 1.3$) with a grid spacing
of $\Delta a = 0.01$ AU and changed also the
eccentricity of the known planet between $0.4 < e < 0.5$ with a gridsize of
$\Delta e = 0.01$. 
%%%%%%%%%%%%%%%%%%%%%%%%
The results of the two methods of analysis of the orbital
behavior are shown in figures~\ref{fig5} and~\ref{fig6}. 
In the first plot we show the results of
the MEM, where two features are immediately
visible: 
1) strong vertical lines due to high order resonances, and 2) unstable orbits due
to high eccentricity and high semimajor axes values (red or yellow
colors). 
The latter feature is easy to understand because closer to the existing planet the
perturbations are larger.   
The two methods complement each other in the information they convey; 
the MEM tells us about the variable distance to the central star
and consequently it is a direct measure of the differential energy flux (insolation) 
on the planet. We can therefore determine where the variation of this distance does not
exceed 50 percent, corresponding to an eccentricity of $e=0.2$. The R\'enyi entropy
is a more sensitive probe of the dynamical character of the orbit, giving us a 
measure of the degree of chaoticity. In particular high order resonance features are 
made very clear using this second method, and we can even see the resonances acting when the
eccentricity of the planet is as low as $e=0.4$ (the bottom of figure~\ref{fig6}). 

As a result for habitability of a terrestrial planet inside the orbit of the
Jovian planet, we find that for the system Gl~777~A there is quite a good chance that 
planets will last long enough in the HZ to acquire the necessary conditions
for life in the region with $a < 1$ AU.

\section{47 Uma}

\begin{figure}[hhh]
\begin{center}
\includegraphics[width=3.4in]{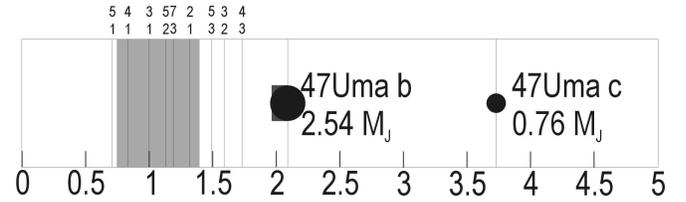}
\caption{Main characteristics of the extrasolar system 47~Uma. Specifications
  like in figure~\ref{fig4}.}
\label{fig7}
\end{center}
\end{figure}

\begin{figure}[hhh]
\begin{center}
\includegraphics[width=3.4in]{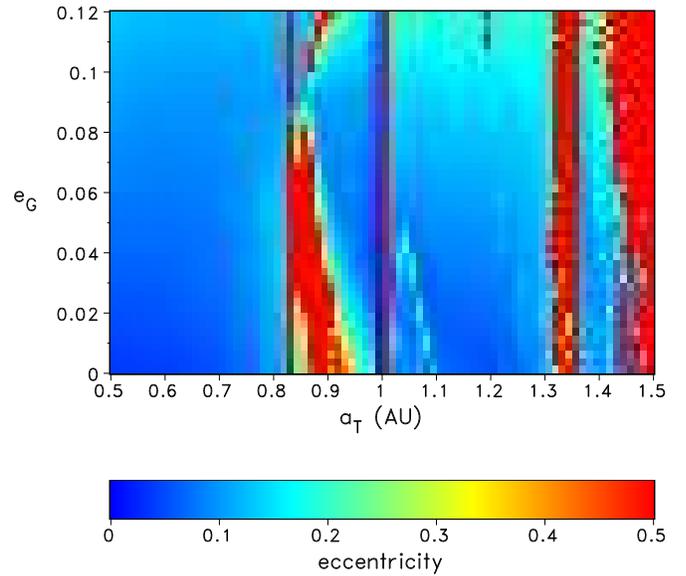}
\caption{Results of the MEM for 47~Uma.}
\label{fig8}
\end{center}
\end{figure}

The first planet in 47 Uma was discovered by Butler \& Marcy~(\cite{Butler96}), and the
second, the outer one, was reported in a paper by Fischer et al.~(\cite{Fischer02}). 
Former studies investigated the zones inside the orbit of the more massive
inner planet (Jones \& Sleep~\cite{Jones02}; Cuntz et al.~\cite{Cuntz03}) which may host 
terrestrial planets. The Jovian planet
has an almost circular orbit with very small errors in the eccentricity. 
Because of the errors for the eccentricity of the inner Jovian planet's orbit,
 we varied its eccentricity 
from $0< e < 0.12$ for our computations. It turned out that the system 
became unstable when $e_{\mathrm{inner}} >0.12$ and consequently also no inner terrestrial planet
would survive. Almost all the MMRs of the fictitious planet with the inner planet are
in the HZ (figure~\ref{fig7}). The respective computations, summarized in table~\ref{tab3}, show that
almost 1/3 of planets located in these resonances would survive with moderate
eccentricities, which can be explained by the small eccentricity of the Jovian
planet. The situation is similar to Jupiter and Saturn but with a scaling factor of approximately 0.4 in $a$.
Also in our Solar System, asteroids are unstable in the outer main belt at and beyond 
the 2:1 resonance, although some islands of stability exist, like the Hilda asteroids in the 3:2 MMR. 
The vertical line in figure~\ref{fig8}, located near
1~AU, indicates unstable orbits that mark the 3:1 resonance.
The thin vertical line of large eccentricities close to a=0.82 AU
corresponds to the 4:1 mean motion resonance of the terrestrial planet with
the inner large planet. We can observe a broad curved ``unstable'' line between
0.8 AU and 0.92 AU, which is caused by secular resonances (=SR) of the
perihelia; this was confirmed by test computations, where we omitted the outer
planet. It is remarkable that these SR act very strong when the initial
eccentricity of the inner planet is close to zero, then it becomes weaker and
disappears for $0.08 < e <  0.1$, but it is again visible for $e>0.1$. It is
interesting to note how the period of the periastron of the giant planet
changes ($4000 \mathrm{ yrs} < \dot \omega < 20000 \mathrm{ yrs}$) with increasing
eccentricity. Between the SR and the 3:1  resonances a small
region of orbits with small enough eccentricities allowing habitability ($e<0.2$)
may survive.
On the outer part of the HZ the very strong 2:1 resonance (broad line close to
1.3 AU) limits the region of stable orbits, but still leaves a large region of stable
orbits between the 3:1 and the 2:1 resonance. 
This is in good agreement with the results of the previously mentioned studies
(e.g. Cuntz et al.~\cite{Cuntz03}),
which did not cover such a large volume of phase space of possible motions in the HZ.
The entropy plot does not show any features other than the ones we can see
from the MEM.

\section{HD 72659}

\begin{figure}[hhh]
\begin{center}
\includegraphics[width=3.4in]{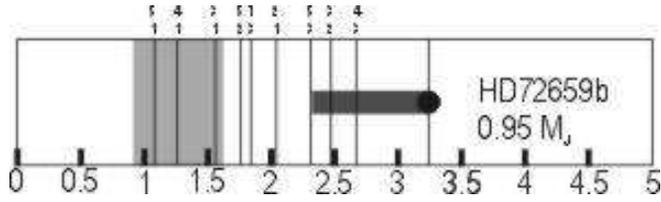}
\caption{Main characteristics of the extrasolar planetary system HD 72659.
 Specifications like in figure~\ref{fig4}.}
\label{fig9}
\end{center}
\end{figure}

The G05 star HD~72659 was found to have a companion from the Keck Precision
Doppler survey (Butler et al.~\cite{Butler02}). The Jovian planet (2.55 $M_{\mathrm{J}}$)
has an orbit with a semimajor axis $a=3.24$ AU and an eccentricity of
$e=0.18$. The MMRs are located from 2.47 AU (3:2) to 1.1081 AU
(5:1); the 5:1, 4:1 and 3:1 are well inside the periastron position of 2.657 AU and
lie in the HZ (around 1 AU, see figure~\ref{fig9}).

The resonances turned out to be stable in more than $40 \%$ of the orbits;
especially the high order resonances close to the HZ are stable in both initial conditions
(periastron and apoastron position). As a consequence we expected that these  
planetary systems may host additional terrestrial planets in stable orbits.

\begin{figure}[hhh]
\begin{center}
\includegraphics[width=3.4in]{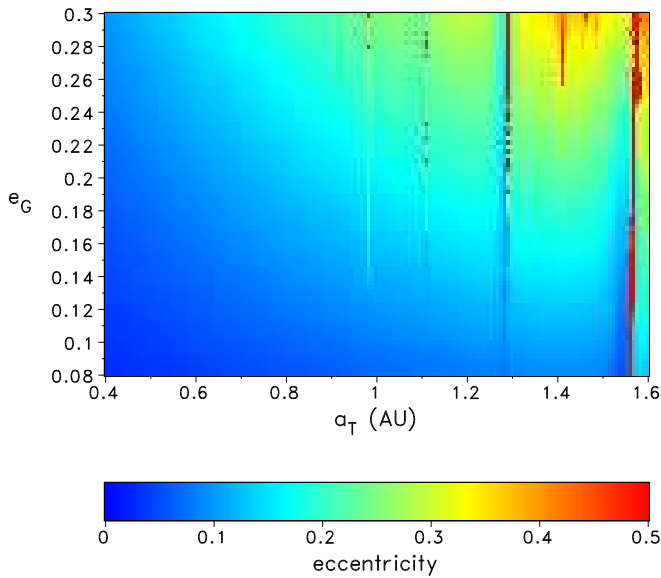}
\caption{Results of the MEM for HD 72659.}
\label{fig10}
\end{center}
\end{figure}

\begin{figure}[hhh]
\begin{center}
\includegraphics[width=3.4in]{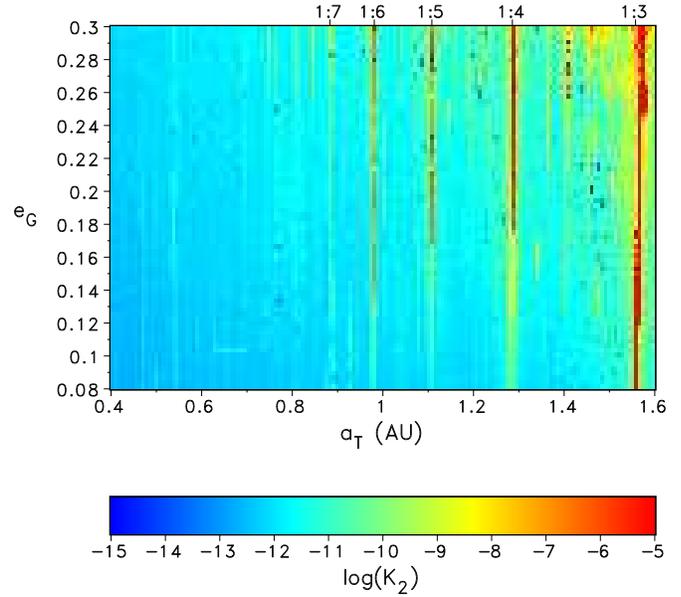}
\caption{Entropy plot for HD 72659.}
\label{fig11}
\end{center}
\end{figure}

Because of the uncertainties  of the observed Jovian planet's eccentricity 
we varied it from 0.08
to 0.30 with a stepsize of 0.22/80 = 0.00275 and chose the initial semimajor
axis of the fictitious terrestrial planet to satisfy $0.4$ AU $< a < 1.2$ AU. The results
are shown in figure~\ref{fig10} (MEM) and figure~\ref{fig11} (entropy plot). 
We can identify quite well in these plots the
resonances up to the 7:1 resonance (only in the entropy plot).
Again one can see that the dynamics of a single orbit can be determined quite
well with this method; it does not only confirm what is depicted in figure~\ref{fig10},
it also shows many more details especially for the motions in resonances. On the
contrary the MEM is the appropriate tool for determining the eccentricity,  which
is -- together with the semimajor
axes -- the crucial parameter for our research of determining planets in habitable zones.

Globally we can see a quite stable HZ in this extrasolar system which allows
planets on orbits with small eccentricities. The strong unstable line close to
1.6~AU corresponds to the 3:1 resonance, while the other resonances, although giving
rise to large perturbations in the eccentricities, are confined to the center
of the resonance.
In figure~\ref{fig10} we can also see that
for the most probable eccentricity value of the Jovian planet ($e=0.18$) all orbits
up to the 3:1 resonance are stable with low eccentricities ($e < 0.2$) and 
as a consequence the HZ could be populated with a terrestrial planet (or even
more planets depending on their masses).

\section{Gl 614}  

\begin{figure}[hhh]
\begin{center}
\includegraphics[width=3.4in]{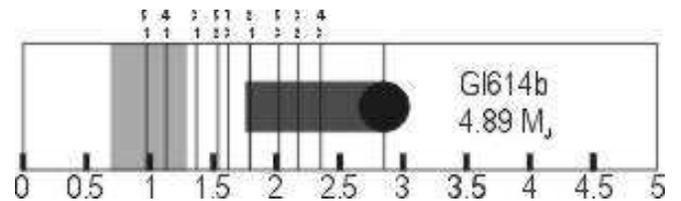}
\caption{Main characteristics of the extrasolar planetary system Gl~614.
Specifications like in figure~\ref{fig4}.}
\label{fig12}
\end{center}
\end{figure}

A planet around the K0V star Gl 614 (=14 Herculis) was discovered by 
Naef et al~(\cite{Naef04}) which is quite
massive ($4.89$ $M_{\mathrm{J}}$) and orbits the central star with a semimajor
axis $a=2.85$ AU
and an eccentricity of $e=0.38$. The computation of orbits in MMRs showed that
none of them are stable up to the 6:1 resonance in apoastron and periastron of
the existing planet (for details see table~\ref{tab3}); this is not surprising when one looks 
at the large mass and the semimajor
axis combined with the large eccentricity: the apoastron of the planet is 1.76 AU. 
The characteristics of this systems are shown in figure~\ref{fig12}. 

\begin{figure}[hhh]
\begin{center}
\includegraphics[width=3.4in]{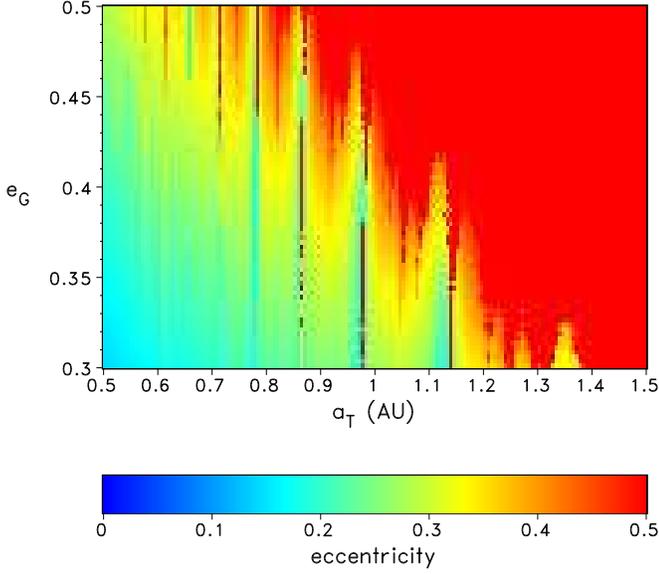}
\caption{Results of the MEM for Gl 614. 
This plot shows that for this system it is very unlikely for a terrestrial planet to 
survive in a stable orbit which also permits liquid water to exist on its surface.}
\label{fig13}
\end{center}
\end{figure}

Looking at the results from the MEM (figure~\ref{fig13}) one can see that for the most
probable value of the eccentricity of the Jovian planet only the orbits with $a<0.7$
stay in the stability and habitability ($e<0.2$) limits within the HZ.

\section{HD 4208}

Using the data of the Keck Precision Velocity Survey, Vogt et al.~(\cite{Vogt02}) 
discovered that the system HD~4208 contains a planet with a slightly smaller mass than
Jupiter on an almost perfectly circular orbit at 1.67 AU. 
The G5V star has a mass almost like our Sun. Our simulations of the
dynamical evolution of terrestrial planets were carried out in the same way as for the other
four systems but with special emphasis on the resonances in this system (see
section 3). Unfortunately, we made an error in the mass of the central star (we took it
to be 14 \% smaller than the one determined by the observers) for the
computations in the HZ (not for the stability in resonances!).
The habitable zone will be shifted outwards only a few percent because of this mass error;
motions in the resonances (in the HZ) will be, for all practical purposes, unaffected.

The MMRs of first, second, third and forth order were investigated for this
system. For a total of 18 resonances in two different positions of
the existing planet (apoastron and periastron) for 8 different values of the
mean anomaly of the fictitious planets we computed the stability (for a
discussion see section~\ref{sec3}).

\begin{figure}[hhh]
\begin{center}
\includegraphics[width=3.4in]{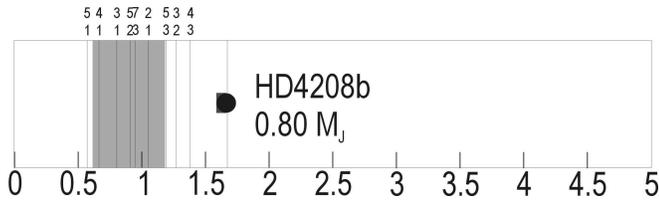}
\caption{Main characteristics of the extrasolar planetary system HD 4208.
Specifications like in figure~\ref{fig4}.}
\label{fig14}
\end{center}
\end{figure}

\begin{figure}[hhh]
\begin{center}
\includegraphics[width=3.4in]{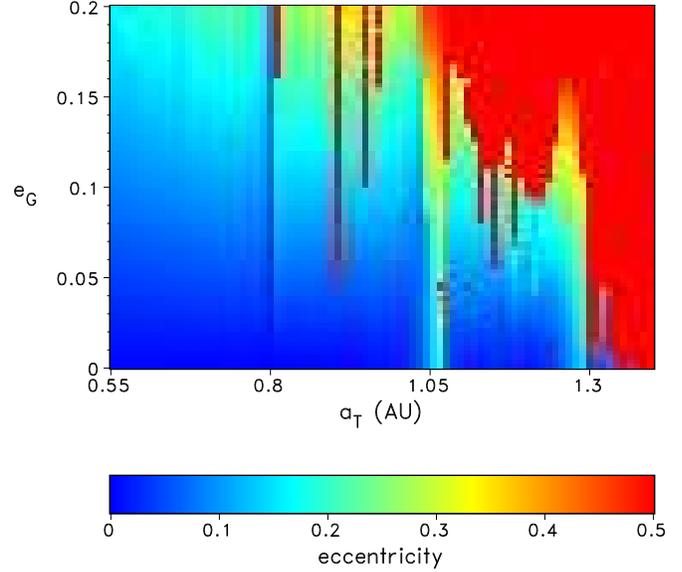}
\caption{Results from the MEM for HD~4208.}
\label{fig15}
\end{center}
\end{figure}

\begin{figure}[hhh]
\begin{center}
\includegraphics[width=3.4in]{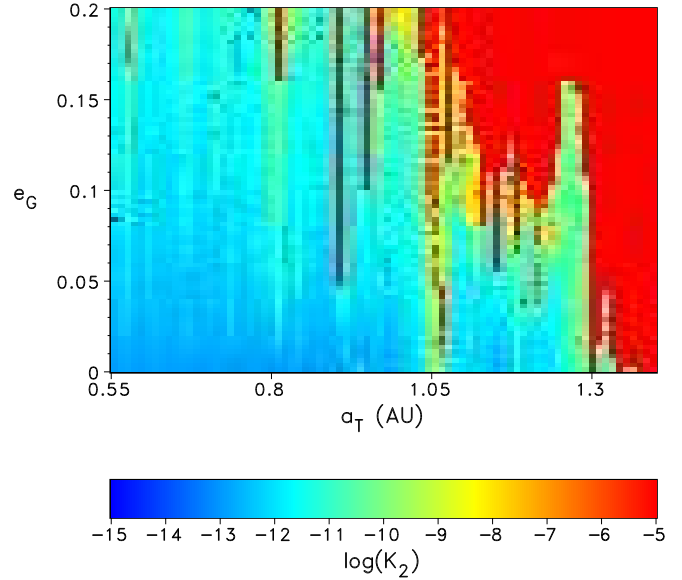}
\caption{Entropy plot for HD~4208.}
\label{fig16}
\end{center}
\end{figure}

We expected unstable motion for this system from a = 1.40 AU on, a limit which we computed
using estimations derived by Wisdom~(\cite{Wisdom80}) and Duncan et al.~(\cite{Duncan89}) for the
onset of  global chaos. We took the mean value between Wisdom's and Duncan's
estimate (namely $\Delta a = 1.27 \mu ^{2/7}$ -- where $\mu$ is the mass ratio
of the primaries -- which leads to a $\Delta a=0.28$
for this system). In the two plots (figures~\ref{fig15} and~\ref{fig16}) one can see that in fact
the orbits are unstable from $a=1.3$ AU on.

From the 2:1 resonance inward, for all initial conditions of the Jovian planet, we can expect  
stable orbits with low enough eccentricities ($e<0.2$) to allow habitable terrestrial planets with 
liquid water, the only exception being the resonant orbits (the vertical lines in both figures).

\section{Conclusions}

We have undertaken a dynamical study of five extrasolar
 planetary systems using extensive numerical experiments to answer 
the question of whether they could host terrestrial like planets in habitable zones. 
For the single--planet systems we used the elliptic restricted three body
 problem, and for the system 47 Uma
the mutual perturbations of the two Jovian planets were taken into account and
therefore the restricted four-body problem served as a dynamical model.
Because of the dependence of the stability of an orbit in a mean motion resonance 
on angular position we have selected 8 different positions for the mean anomaly of the planet
with a step of $45^\mathrm{\circ}$ in the  apoastron and
the periastron position of the Jovian planet.

The characteristics of each system dictated the initial conditions, chosen in
a fine 80 by 80 grid within the habitable zone, from which the orbits were
computed using a robust numerical method (Lie--series integration) 
for 1 million years. 
The grid of the initial conditions of the fictitious terrestrial planets was chosen
to cover the whole habitable zone of the system, and also
to model the uncertainties in the elements of the observed planet(s). 
The stability of orbits was assessed with two methods, namely
the computation of the R{\'e}nyi entropy as measure of the chaoticity of an orbit
and the determination of the maximum eccentricity of the orbit
of a fictitious planet during its orbital evolution of 1 million years.

We can say that our computations for such a fine grid, taking into account also
the essential role of the MMRs, lead to a deeper insight concerning the
dynamics of the five systems which we studied.  
We also give the percentage of orbits which survived in the paper (=MT) of Menou \& 
Tabachnik~(\cite{Menou03}) where they
investigated all known extrasolar planetary systems with respect to possible 
additional terrestrial planets. We note that a direct comparison of MT with
the percentages of 'our' survivors is not useful here because of the different 
approaches used; we have emphasized the role of the MMRs and neglected
possible inclinations. However, we know that terrestrial planets will form
within a
protoplanetary disk thus staying with small orbital inclinations (Richardson
et al.~\cite{Richardson00}; Lissauer~\cite{Lissauer93}); additionally  in a recent publication
(Dvorak et al.~\cite{Dvorak03a}) it was shown that the inclinations of the fictitious planets up
to 15 degrees do not change the stability of orbits in the HZ.
The results for possible stable orbits may be summarized
as follows:

\begin{itemize}

\item In the system {\bf Gl~777~A} the stability zone for the motion of terrestrial
planets is well inside of the HZ and suggests any planets residing there
will survive for a sufficiently long time (in MT, 86.8 \% of the orbits were found to be stable).

\item  In the system {\bf 47 Uma} we can also say that there is a good chance for planets
 to move inside the HZ with small eccentricities between the main resonances;
this is a result which 
is consistent with others but not with MT where only 28 \% remained.

\item {\bf HD 72659} turned out to be a very good candidate for hosting planets in
  the HZ; again this does not confirm the results of MT -- they found that only 40.2
  \% of the orbits were stable.

\item    The results of the computations for {\bf Gl 614} show that  it is
  very unlikely that there is an
  additional planet moving in the HZ -- these results are consistent 
  with MT (9.2 \% stable orbits)

\item For {\bf HD 4208}  we can say that in the HZ there is enough room left for
terrestrial planets and that they could survive for sufficiently long time;
these results are more or less consistent with those of MT where 50.2 \% of
the orbits were stable.
  
\end{itemize}

New observational possibilities provided by missions like COROT, DARWIN or the
TPF make the first search for terrestrial exoplanets
seem possible in the next decades. In an ESA study the goal of the
missions is summarized as follows: ''To detect and study Earth--type planets
and characterize them as possible abodes of life''. In this sense dynamical
studies like the one we present here should help to define promising
targets for observations.

\acknowledgements

We all deeply thank the Helmholtz Institute for hosting the
summerschool 2003 and for providing the necessary financial support. 
We also acknowledge the generosity of the Max Planck Albert Einstein Institut
where we were able to perform these massive computations on their new high 
performance cluster, PEYOTE.
Special thanks go to Profs. Kurtz and Ruediger, the directors of the HISP in
Potsdam for their hospitality.

\appendix

\section{Recurrence Plots}
\label{ApA}
\subsection{Introduction}
In his seminal paper Henri Poincar\'e~(\cite{poincare}) introduced the
concept of recurrences in phase space, when he studied the stability of the
solar system. Recurrence Plots (RPs) were introduced to visualize the
recurrences of trajectories of dynamical systems in phase space (Eckmann et
al.~\cite{eckmann87}). These plots have proved to be rather useful in the
analysis of time series, as they give a first impression about the behavior of
the system under study. However, in order to go beyond mere visual inspection,
which depends on the person doing the analysis and hence is always to some
extent subjective, measures that quantify the
structures found in RPs were introduced (Webber \& Zbilut~\cite{Webber}). 
This quantification of RPs has found numerous
applications in many fields, such as Geology (Marwan et
al.~\cite{norbert_geo}), Physiology (Marwan et al.~\cite{norbert_heart};
Zbilut et al.~\cite{zbilut_cardiac}), Climatology~(Marwan \& Kurths et
al.~\cite{norbert_crps}), etc. Especially, it has been shown that it is
possible to estimate dynamical invariants based on RPs, like the R\'enyi
entropy and the correlation dimension (Thiel et al.~\cite{thiel_and}). The
estimation of these invariants by means of RPs has some advantages with
respect to other algorithms usually applied to the analysis of time series.
The method of RPs avoids, for example, the problem of the choice of embedding
parameters, and it can also be applied to non-stationary data~(Thiel et
al.~\cite{thiel_chaos}). We concentrate on the estimation of the R\'enyi
entropy, as it can be interpreted as the inverse of the prediction time 
(or Lyapunov timescale), i.e., it can be used to measure the prediction 
horizon of the system.

\subsection{Definition of Recurrence Plots and Examples} 

RPs were introduced to simply visualize the behavior of trajectories in phase space~(Eckmann et
al.~\cite{eckmann87}). Suppose we have a dynamical system represented by the
trajectory $\{\vec x_i\}$ for $i=1,\ldots,N$ in a $d$-dimensional phase
space. Then we compute the recurrence matrix 

\begin{equation}\label{eq1}
R_{i,\,j} = \Theta(\varepsilon-\left||\vec x_{i} - \vec x_{j}\right||), \quad \, 
i, j=1\dots N,
\end{equation}

where $\varepsilon$ is a predefined threshold and $\Theta(\cdot)$ is the
Heaviside function\footnote{The norm used in Eq.~\ref{eq1} is in principle
  arbitrary. For theoretical reasons, it is preferable to use the maximum
  norm.}. The graphical representation of $R_{i,\,j}$, called a Recurrence Plot,
is obtained by encoding the value one for a ''black'' point and zero for a ''white''
point. The recurrence rate $RR$ is defined as the percentage of black points
in the RP, i.e.  
\[
RR= \frac{1}{N^2}\sum\limits_{i,j=1}^{N}R_{i,j}
\]
\begin{figure*}
\begin{center}
\includegraphics[width=5in]{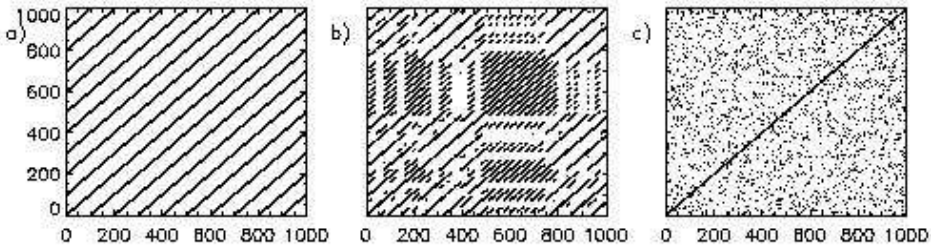}
\caption{Prototypical examples of RPs: a) RP of a sine function, b) RP of the
  R\"ossler system in chaotic regime, c) RP of white noise.} 
\label{rps_ex}
\end{center}
\end{figure*}

Figure~\ref{rps_ex} a) shows the RP of a sine function, i.e a circle in phase
space. The plot is characterized by non-interrupted diagonal
lines. Figure~\ref{rps_ex} b) is the RP of the R\"ossler system in a chaotic
regime. In this case the predominant structure are diagonal lines, which are
interrupted. Figure~\ref{rps_ex} c) represents the RP of white noise. It is
homogeneous with mainly single points, indicating a stochastic system, as the
state at time $t+1$ is independent of the one at $t$. From these plots, we see
that there is a certain connection between the length of diagonal lines and
the ratio of determinism or predictability inherent to the system. This
connection can be explained as follows: suppose that the states at $t=i$ and
$t=j$ are neighboring, i.e. $R_{i,j}=1$. If the system behaves predictably,
then similar situations lead to a similar future, i.e. the probability for
$R_{i+1,j+1}=1$ is high. For perfectly predictable systems, this leads to
infinitely long diagonal lines (like in the RP of the sine function). In
contrast, if the system is stochastic, the probability for $R_{i+1,j+1}=1$ is
small, i.e., we find single points or short lines. If the system is chaotic,
initially neighboring states diverge exponentially. The faster the
divergence, i.e. the higher the Lyapunov exponent, the shorter the
diagonals.
In the next section we develop this observation and explain formally the 
relationship between the length of the diagonal lines in the RP and the
predictability of the system. 

\subsection{Estimation of the R\'enyi entropy based on the
  RP\label{main_formula}}

In this section we recall first the definition of the R\'enyi entropy of
second order and then present the mathematical relation between this
information measure and the RPs.
Let us consider a trajectory $\vec x(t)$ in a bounded $d$-dimensional phase
space and the state of the system is measured at time intervals $\tau$. Let
$\{1,2,...,M(\varepsilon)\}$ be a partition of the attractor in boxes of size
$\varepsilon$. Then $p(i_1,...,i_l)$ denotes the joint probability that $\vec
x(t=\tau)$ is in the box $i_1$, $\vec x(t=2\tau)$ is in the box $i_2$, ...,
and $\vec x(t=l\tau)$ is in the box $i_l$. The order--2 R\'enyi entropy
(R\'enyi~\cite{renyi}; Grassberger~\cite{Grassb0}) is then defined as  
\begin{equation}
K_2=-\lim_{\tau\to 0}\lim_{\varepsilon \to 0}\lim_{l \to \infty}\frac{1}{l\tau}\ln \sum_{i_1,...,i_l} p^2(i_1,\dots,i_l).
\end{equation}
Roughly speaking, this measure is directly related to the number of possible
trajectories that the system can take for $l$ time steps in the future. If
the system is perfectly deterministic, in the classical sense, there is only
one possibility for the trajectory to evolve and hence $K_2=0$. In contrast,
one can easily show that for purely stochastic systems, the number of possible
future trajectories increases to infinity so fast, that $K_2 \to
\infty$. Chaotic systems are characterized by a finite value of $K_2$, as they
belong to a category between pure deterministic and pure stochastic
systems. Also in this case the number of possible trajectories diverges but
not as fast as in the stochastic case. The inverse of $K_2$ has units of time
and can be interpreted as the mean prediction horizon/time of the system.

\begin{figure*}
\begin{center}
\includegraphics[width=5in]{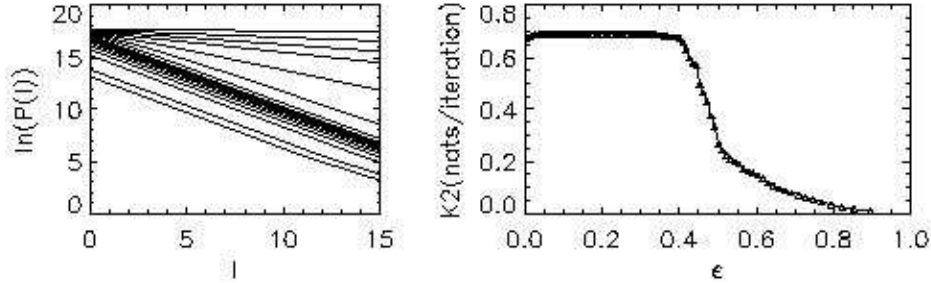}
\caption{Left panel: number of diagonal lines of at least length $l$
  versus $l$ in the RP of the Bernoulli map for different values of the
  threshold $\varepsilon$. The mean slope of the curves is equal to
  $0.6917$. Right panel: R\'enyi entropy vs. $\varepsilon$ for the Bernoulli
  map.} 
\label{bern}
\end{center}
\end{figure*}

In the paper of Thiel et al.~(\cite{thiel_and}) it was shown that there exist
a direct relationship between $K_2$ and RPs, and it is as follows 
\begin{equation}
\ln P^c_{\varepsilon}(l) \sim \varepsilon^{D_2}\exp(-\hat{K}_2(\varepsilon)\tau),
\end{equation}
where $P^c_{\varepsilon}(l)$ is the cumulative distribution of diagonal lines
in the RP, i.e. the probability of finding a diagonal in the RP of at least
length $l$ \footnote {Fractions of longer lines are also counted: e.g. a line
  of length 5 counts as 5 lines of length 1, 4 lines of length 2, and so
  on.}. $D_2$ is the correlation dimension of the system under consideration
(Grassberger \& Procaccia 1983). Therefore, if we represent
$P^c_{\varepsilon}(l)$ in a logarithmic scale versus $l$ we should obtain a
straight line with slope $-\hat{K}_2(\varepsilon)\tau$ for large $l$'s, which
is independent of $\varepsilon$ for a rather large range in
$\varepsilon$. This is shown in the left panel of figure~\ref{bern} for the
chaotic Bernoulli map $x_{n+1}=2 x_n mod(1)$. One obtains $\hat K_2=0.6733$
for 10000 data points, in good accordance with the values found in the
literature~(Ott~\cite{ott}).

Finally one represents the slope of the curves for large $l$ in dependence on 
$\varepsilon$, where a plateau is found for chaotic systems. The value of 
this plateau determines $\hat{K}_2$ (Figure~\ref{bern}, right panel). If the system is not chaotic, one has 
to consider the value of the slope for a sufficiently small value of $\varepsilon$.  

\subsection{Automatization of the Algorithm}

To compute the stability diagrams presented in this paper, which consist of
about 13000 grid points, the estimation of $K_2$ had to be automated. The
crucial point in the automatization is the estimation of the scaling region of
$\ln P^c_{\varepsilon}(l)$ vs. $l$ and the plateau in $K_2(\varepsilon)$
vs. $\varepsilon$. In both cases we applied a cluster dissection algorithm
(Spaeth~\cite{spaeth}). The algorithm divides the set of points into distinct
clusters and performs  a linear regression in each cluster. Then the sum of
all square residuals is minimized in order to determine the scaling
region and the plateau. To find both regions automatically, we used the
following parameters: 
\begin{itemize}
\item
We considered only diagonal lines up to length $l_{max}=400$. 
Longer lines were excluded because of finite size effects.
\item
We considered only values of $P^c_{\varepsilon}(l)$ with 
$P^c_{\varepsilon}(l)> 500$ for the same reason as in the last item.
\item
We used 40 different values for $\varepsilon$, corresponding to 40 equally
spaced recurrence rates $RR$ between 1\% and 95\%, to have a well defined
plateau in $K_2(\varepsilon)$ vs. $\varepsilon$. 
\item
We used 10000 data points of each simulated trajectory. The more data points
one uses, the more pronounced the scaling regions. Note that the computation
time increases approximately with $N^2$. 
\item
For the applied cluster dissection algorithm we had to specify the number of
clusters in each run. For the detection of the scaling region in $\ln
P^c_{\varepsilon}(l)$ vs. $l$, we chose 2 different clusters and used the
slope of the largest cluster. For the detection of the plateau in
$K_2(\varepsilon)$ vs. $\varepsilon$, we chose 3 clusters and used the value
of the cluster with the minimum absolute slope. 
\end{itemize}
These choices have proven to be the most appropriate ones for the estimation
of the scaling regions. All these parameters are defaults of the computer
program. The only needed input is the file with the trajectory data.

\section{The Lie-series Method}
\label{ApB}
\subsection{General Properties of the Lie-series}
Gr\"obner~(\cite{Groebner67}) defined the Lie-operator $D$ as follows:
 
\begin{equation}
D=\theta_1(z)\frac{\partial}{\partial z_1} + \theta_2(z)\frac{\partial}{\partial z_2}
+ \dots + \theta_n(z)\frac{\partial}{\partial z_n}
\label{def}
\end{equation}

$D$ is a linear differential operator; the point $z=(z_1,z_2,\dots,z_n)$ lies
in the n-dimensional z-space, the functions $\theta_i(z)$ are holomorphic
within a certain domain $G$, e.g. they can be expanded in a converging power
series. Let the function $f(z)$ be holomorphic in the same region as
$\theta_i(z)$. Then $D$ can be applied to $f(z)$:

\begin{equation}
Df=\theta_1(z)\frac{\partial f}{\partial z_1} + \theta_2(z)\frac{\partial f}{\partial z_2}
+ \dots + \theta_n(z)\frac{\partial f}{\partial z_n}
\end{equation}

If we proceed applying $D$ to $f$ we get

\begin{eqnarray}
D^2f&=&D(Df) \nonumber \\
\vdots&& \nonumber \\
D^nf&=&D(D^{n-1}f) \nonumber
\end{eqnarray}
The {\bf Lie-series} will be defined in the following way;

\begin{displaymath}
L(z,t)=\sum_{\nu=0}^{\infty} \frac{t^\nu}{\nu!} D^\nu f(z) = f(z)+  tDf(z) +
\frac{t^2}{2!}D^2f(z)+\dots 
\end{displaymath}
Because we can write the Taylor-expansion of the exponential function
\begin{equation}
e^{tD} = 1 + tD^1 + \frac{t^2}{2!}D^2+\frac{t^3}{3!}D^3+\dots
\end{equation}
$L(z,t)$ can be written in the symbolic form
\begin{equation}
L(z,t)=e^{tD}f(z)
\end{equation}

The convergence proof of $L(z,t)$ is given in detail in Gr\"obner~(\cite{Groebner67}).

One of the most useful properties of Lie-series is the {\em Exchange Theorem}:

{\sf Let $F(z)$ be a holomorphic function in the neighborhood of
  $(z_1,z_2,\dots,z_n)$ where the corresponding power series expansion
  converges at the point $(Z_1,Z_2,\dots,Z_n)$; then we have:}

\begin{equation}
F(Z)= \sum_{\nu=0}^{\infty} \frac{t^\nu}{\nu!} D^\nu F(Z)
\end{equation}

{\sf or}

\begin{equation}
F(e^{tD})z=e^{tD}F(z)
\end{equation}
Making use of it we can demonstrate how Lie-series solve DEs. Let us give the
system of DEs:

\begin{equation}
\frac{dz_i}{dt}=\theta_i(z)
\label{des}
\end{equation}

with $(z_1,z_2,\dots,z_n)$. The solution of~\ref{des} can be written as

\begin{equation}
z_i=e^{tD}\xi_i
\label{res}
\end{equation}

where the $\xi_i$ are the initial conditions $z_i(t=0)$ and D is the
Lie-operator as defined in~\ref{def}. In order to prove~\ref{res} we
differentiate it with respect to the time $t$:

\begin{equation}
\frac{dz_i}{dt}=De^{tD}\xi_i=e^{tD}D\xi_i
\end{equation}

Because of

\begin{equation}
D\xi_i=\theta_i(\xi_i)
\end{equation}

we obtain the following result which turns out to be the original DE~\ref{des}:

\begin{equation}
\frac{dz_i}{dt}=e^{tD}\theta_i(\xi_i)=\theta_i(e^{tD}\xi_i)=\theta_i(z_i)
\end{equation}

\subsection{A simple example}

To demonstrate the principle of the Lie-integration, we will show how one
proceeds in the simple case of the harmonic oscillator, a $2^{nd}$ order DE
where the solution is known:

\begin{equation}
\frac{d^2x}{dt^2}+\alpha^2 x = 0 
\end{equation}

The first step consists in separating into two $1^{st}$ order DEs such that
\begin{displaymath}
\frac{dx}{dt}=y=\theta_1(x,y)
\end{displaymath}
\begin{displaymath}
\frac{dy}{dt}=-\alpha^2x=\theta_2(x,y) 
\end{displaymath}
with the initial conditions $z(t=0)=\xi$ and $y(t=0)=\eta$. With this notation
we find the Lie-operator of the form
\begin{equation}
D=\theta_1\frac{\partial}{\partial \xi} + \theta_2\frac{\partial}{\partial
  \eta} = \eta\frac{\partial}{\partial \xi} -
  \alpha^2\xi\frac{\partial}{\partial \eta} 
\end{equation}

The solution can now be written as a Lie-series
\begin{equation}
x=e^{\tau D}\xi \hskip0.7cm \mathrm{and} \hskip0.7cm y=e^{\tau D}\eta
\end{equation}
where $t-t_0=\tau$ and for $\tau=0$ this are the initial conditions. Being
aware of the symbolic development of $e^{\tau D}$ we can compute the first
terms:
\begin{displaymath}
D^1\xi=\eta=\theta_1
\end{displaymath}
\begin{displaymath}
D^2\xi=D\eta=-\alpha^2\xi=\theta_2
\end{displaymath}
\begin{displaymath}
D^3\xi=-\alpha^2 D\xi=-\alpha^2 \eta
\end{displaymath}
\begin{displaymath}
D^4\xi=-\alpha^2 D\eta=\alpha^4 \xi
\end{displaymath}
\begin{displaymath}
D^5\xi=\alpha^4 D\xi=\alpha^4 \eta
\end{displaymath}
\begin{displaymath}
D^6\xi=\alpha^4 D\eta=-\alpha^6 \xi
\end{displaymath}
\begin{displaymath}
\vdots
\end{displaymath}

For the Lie-terms {\sf even} respectively {\sf odd} in the order consequently
one can find
\begin{displaymath}
D^{2n}\xi=(-1)^n\alpha^{2n}\xi
\end{displaymath}
\begin{displaymath}
D^{2n+1}\xi=(-1)^n\alpha^{2n}\eta \,
\end{displaymath}
which leads to the solution for $z$:
\begin{displaymath}
z=\xi+\tau\eta-\frac{\tau^2}{2!}\alpha^2\xi-\frac{\tau^3}{3!}\alpha^2\eta +
\frac{\tau^4}{4!}\alpha^4\xi \dots
\end{displaymath}

Finally we get after the factorization of $\xi$ and of $\eta$
\begin{displaymath}
z=\xi\left(1-\frac{\tau^2}{2!}\alpha^2+\frac{\tau^4}{4!}\alpha^4-\frac{\tau^6}{6!}\alpha^6+\dots\right)+
\end{displaymath}
\begin{displaymath}
\frac{\eta}{\alpha}\left(\tau\alpha-\frac{\tau^3}{3!}\alpha^3+\frac{\tau^5}{5!}\alpha^5-\frac{\tau^7}{7!}\alpha^7+\dots\right) \,
\end{displaymath}
which is exactly the known solution of the harmonic oscillator:
\begin{displaymath}
z(t)=\xi\cos\alpha\tau + \frac{\eta}{\alpha}\sin{\alpha\tau} 
\end{displaymath}

\end{document}